# Introducing flexible perovskites to the IoT world using photovoltaic-powered wireless tags

Sai Nithin R. Kantareddy[1,2], Rahul Bhattacharya[1], Sanjay E. Sarma[1]

[1]Auto-ID Labs, Department of Mechanical Engineering, Massachusetts Institute of Technology, Cambridge, USA

Ian Mathews[2], Janak Thapa[2], Liu Zhe[2], Shijing Sun[2], Ian Marius Peters[2], Tonio Buonassisi[2]

[2]MIT PV Lab, Department of Mechanical Engineering, Massachusetts Institute of Technology, Cambridge, USA

*Abstract*—**Billions of everyday objects could become part of the Internet of Things (IoT) by augmentation with low-cost, long-range, maintenance-free wireless sensors. Radio Frequency Identification (RFID) is a low-cost wireless technology that could enable this vision, but it is constrained by short communication range and lack of sufficient energy available to power auxiliary electronics and sensors. Here, we explore the use of flexible perovskite photovoltaic cells to provide external power to semi-passive RFID tags to increase range and energy availability for external electronics such as microcontrollers and digital sensors. Perovskites are intriguing materials that hold the possibility to develop high-performance, low-cost, optically tunable (to absorb different light spectra), and flexible light energy harvesters. Our prototype perovskite photovoltaic cells on plastic substrates have an efficiency of 13% and a voltage of 0.88 V at maximum power under standard testing conditions. We built prototypes of RFID sensors powered with these flexible photovoltaic cells to demonstrate real-world applications. Our evaluation of the prototypes suggests that: i) flexible PV cells are durable up to a bending radius of 5 mm with only a 20 % drop in relative efficiency; ii) RFID communication range increased by 5x, and meets the energy needs (10-350 µW) to enable self-powered wireless sensors; iii) perovskite powered wireless sensors enable many battery-less sensing applications (e.g., perishable good monitoring, warehouse automation)**

*Index Terms*—**Internet of Things (IoT), Radio Frequency Identification (RFID), Energy Harvesters, Photovoltaics, Sensors**

## I. Introduction

The prospect of augmenting billions of everyday objects with low-cost wireless sensors to create a network of connected objects has driven research in the Internet of Things (IoT) area for a number of years [1]–[3]. A network of connected objects enables us to acquire rich environmental information, and build powerful data-driven applications for places where we live and work [4]–[7]. Augmentation of billions of objects requires us to use low-cost wireless sensors that are passive without needing battery replacements or access to direct power. Radio Frequency Identification (RFID) provides a scalable and energy efficient way to create such passive sensors without requiring an active radio and consuming only a few µW of power [8]–[10].

The use of traditional passive tags is constrained by their short communication range and lack of on-board power to run auxiliary electronics such as external sensors. The short communication range of tags implies that a dense setup of antennas and RFID readers is needed to provide the required coverage — thereby increasing infrastructure and deployment costs. By providing more power locally, using energy harvesters, we can increase the communication range and power auxiliary electronics.

Photovoltaic (PV) energy harvesters can be used to supply energy to RFID tags [11], [12] to overcome this power constraint. Generally, traditional PV technology such as (Si) PV is the preferred choice for PV energy harvesters. Si PV cells are suitable for large-scale outdoor energy harvesting because of their high performance and low price point. However, they are rigid, Silicon heavy, difficult to optimize for artificial lighting, not printable (for potential IoT applications that requires printing of smart labels)), and are expensive to manufacture at small scale. In this study, we introduce flexible perovskite PV for win in IoT. We discuss applications to show how perovskite PV technology can meet the needs of low-power IoT devices. Our earlier publications ([8], [13]) focused on standard perovskite PV cells on glass substrates, resulting in mechanically rigid devices. In this study, we demonstrate fully flexible perovskite PV cells that are appealing to integrate with consumer products. Examples of flexible PV-powered RFID tags in the available literature [12], [14], [15] use amorphous Silicon (a-Si) PV cells. Perovskite PV is superior to a-Si PV technology in terms of efficiency, optical tunability (to absorb different light spectra), and printability (more readily available to integrate with RFID tags at low costs)[16].

The authors acknowledge the funding sources for this work. S.N.R.K. received funding from the Legatum Center at MIT. I.M. received funding from the European Union's Horizon 2020 research and innovation programme under the Marie Skłodowska-Curie grant agreement no. 746516. S.S. and J.T. received funding through C2C International Collaboration on Advanced Photovoltaics grant. I.M.P. received funding from the DOE-NSF ERF for Quantum Energy and Sustainable Solar Technologies (QESST).

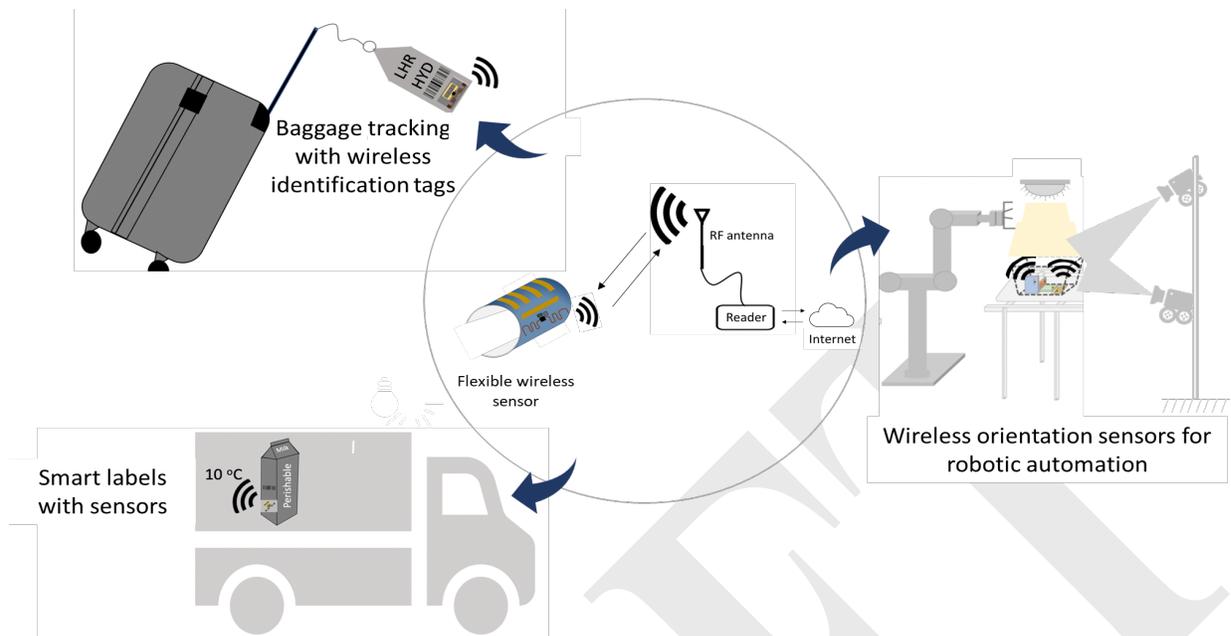

Figure 1: Illustration of plastic perovskite-powered wireless sensors and tag applications: potential baggage tags augmented with perovskite-powered RFID tags, smart sensor labels to monitor perishable goods in supply chains, and inexpensive orientation detection labels to augment camera data with sensor data in warehouse automation

There are currently several energy harvesting technologies for powering IoT devices, but perovskite PV has unique advantages in its ability to be customized. Table 1 summarizes the pros and cons of different energy harvesting technologies. PV has the highest energy harvesting potential and can be used in a variety of environments (e.g. outdoors, indoors, cold/hot, noise-less, and non-RF environments). Flexible, low-cost, optically tunable, and color customizable perovskite photovoltaics have many synergies with low-cost, low-energy, and scalable RFID technology, leading to realizing potentially long-range wireless sensors for just a few cents. Figure 1 illustrates potential applications of flexible perovskite-powered wireless sensors and tags for real-time baggage tracking[1], sensing of perishable items in supply chains using smart labels, and improving robotic automation in warehouses with sensor-camera data fusion using orientation information obtained from low-cost sensor labels.

Main contributions of this work are:
a) Developed flexible perovskite PV-powered wireless sensors that are suitable to create conformal sensor labels for consumer products.
b) Tested and evaluated the performance of flexible perovskite PV for powering RFID tags to increase the range 5 times and provide energy for auxiliary electronics.
c) Explored innovative IoT applications using self-powered wireless tags in industry relevant use cases: Demonstrated prototypes of temperature, orientation and activity recognition sensors.

In the rest of this paper, we first report the device architecture and fabrication of flexible perovskite PV cells and their integration with RFID tags, in Section II. In Section III, we present results from testing the perovskite photovoltaics. In Section IV, we present results from testing the prototypes of self-powered wireless temperature, orientation and activity recognition sensors combining perovskite PV and RFID. We also present how these cells can be used in real-world applications such as warehouse automation. Finally, we present an outlook towards further improving the applicability of perovskite PV for IoT applications by focusing on improving manufacturing, stability, and reduced environmental risk.

---

[1] https://news.delta.com/delta-introduces-innovative-baggage-tracking-process

Table 1 Comparison of perovskite PV with other potential external energy sources to power RFID tags

| External energy source | Pros | Cons | Available energy density [51], [52], [53] |
|---|---|---|---|
| Traditional Si PV [54] | Proven long life times (> 10yrs), high efficiency, market ready, low cost, and high energy density source | Low performance under low-light conditions, difficult to optimize to different lighting conditions, rigid, heavy, and no manufacturing synergies with RFID | 100 mW/cm$^2$ (Outdoor) 100 µW/cm$^2$ (Indoors —underneath light fixtures) |
| Flexible thin-film Si PV | Commercially available dominant indoor photovoltaics technology currently. Thin, lightweight, and flexible modules. | • Low efficiency under low-light conditions<br>• Thin-film Si solar cells are also more expensive and heavier than perovskite PV.<br>• Si also uses high-temperature processing<br>• Fixed bandgap implies not tunable for absorption under different lighting conditions. | 100 mW/cm$^2$ (Outdoor) 100 µW/cm$^2$ (Indoors —underneath light fixtures) |
| Perovskite PV [8], [13], [55] | Mechanically flexible, high efficiency, manufacturable on plastic inlays, optically tunable, printable, low-temperature processable, and potentially exploit other applications of perovskites (e.g. energy storage and tunable antennas) | Stability, standard design contains Pb, and currently not on the market | 100 mW/cm$^2$ (Outdoor) 100 µW/cm$^2$ (Indoors—underneath light fixtures) |
| RF energy harvesters [17] | Mechanically flexible, printable antennas, low cost, and increasing wireless coverage | Low energy density, detunes due to environmental noises, and transmission power limits | 0.015 µW/cm$^2$ (WiFi) 0.03 µW/cm$^2$ (GSM) |
| Thermoelectric generators [18], [19] | Thin form factor, semi-transparent, and high-energy density source | Non-ubiquitous usage and relatively higher cost | 100 µW/cm$^2$ (Human skin to ambient temperature differential) |
| Vibration energy harvesters [20] | Potential to harvest environmental vibrations, potential to embed (e.g. in roads), and niche applications (e.g. car tires) | Low energy density and relatively higher cost | 800 µW/cm$^3$ (Machine motion) |

## II. RFID Tags Flexible and Perovskite Photovoltaics

### A. RFID communication range and power consumption

Passive RFID tags communicate by backscattering RF signals from the RFID reader. Therefore, an RFID tag does not require any large electronics or an active radio to generate RF signals. An RFID tag only consists of an antenna and an Integrated Chip (IC). The passive tag communication is forward communication link limited due to the limit on the reader transmit power. The read range can be estimated by applying the link budget equation (1) as explained in [21].

$$P_{IC} = \frac{P_T G_{tag} G_{reader}}{4\pi d^2} * \frac{\lambda^2}{4\pi} * \tau \quad [1]$$

Therefore,

$$d = \sqrt{\frac{P_T G_{tag} G_{reader} \tau}{P_{IC}}} * \frac{\lambda}{4\pi}$$

$$= constant * \sqrt{\frac{P_T}{P_{IC}}} \quad [2]$$

where $P_{IC}$: received power at the IC, $P_T$: reader transmit power, $d$: distance between the reader and the tag, $\lambda$: RF signal wavelength, $G_{tag}$, $G_{reader}$: tag and reader antenna gains, respectively, and $\tau$: transmission coefficient

If the strength of the incident RF signals ($P_{IC}$) is higher than the power required to turn on the IC (state-of-the-art IC sensitivity is -23 dBm), the IC can power up and backscatter the modulated RF signals for communication. The IC encodes the information by modulating an internal impedance according to the standard EPC Gen 2 protocol [22]. However, if a tag is located far away from the reader, the incident RF signals are weak and cannot power up the IC. The reader transmit power ($P_T$) is limited to a maximum of 1 W according to the guidelines set by the Federal Communications Commission (FCC), therefore, $P_T$ cannot be unrestrictedly increased to increase the communication range ($d$). However, passive RFID tags can be converted into semi-passive RFID tags by providing energy from external sources — thereby reducing the IC's dependence on incident RF signal strength. In the semi-passive case, the IC takes power (around 10 µW depending on operation) from an external source instead of harvesting the incident RF signals to power itself. This reduced dependence on forward link signal strength increases the communication range by a few meters. As confirmed in our previous studies, the range can be improved 5x using external power [8], [11].

Read ranges of our prototype passive and perovskite-powered RFID tags are experimentally measured using a precise RFID testing setup (Voyantic's Tagformance). The results show that the communication range of perovskite-powered RFID tag reached up to 5m compared to only ~1m range obtained with passive devices (see Figure 2). Read range can be further extended by designing antennas with better impedance matching on the tags, for example, antenna designs that considers background dielectrics of perovskite PV cells. Ranges over 10s of m are theoretically achievable as estimated in [11].

Unlike the passive RFID ICs, semi-passive ICs allow external power input. Figure 2 shows an illustration of a photovoltaic powered tag with auxiliary electronics. In the simplest architecture, the tag consists of a "squiggle-shaped antenna" connected to an RFID IC and a flexible PV module connected to the voltage pins of the IC. We used the commercially available EM 4325[2] and Farsen's Rocky 100 [3] ICs to build prototypes in this study. All components can be assembled on a plastic substrate, realizing fully flexible photovoltaic-powered wireless tags. RFID tag operates in 902-928 MHz in the United States. For the purposes of evaluation, we made use of a circularly polarized antenna (with a gain of 8.5 dBi) connected to an Impinj Speedway RFID reader.

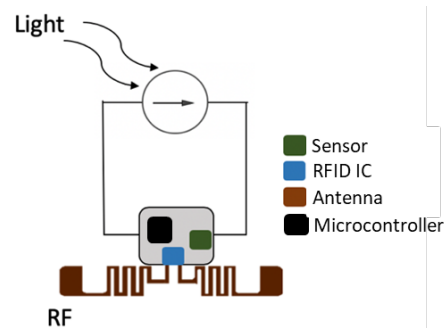

Figure 2: A schematic of perovskite photovoltaic-powered RFID tag with auxiliary electronics [1]

### B. Background on perovskite PV

A perovskite is any material sharing an $ABX_3$ crystal structure. A and B are cations of different sizes and X is an anion. Perovskite materials are shown to exhibit varied and tunable properties, depending on the A, B and X ions, such as super-conductance [23], magnetoresistance [24], dielectric [25], and photovoltaic [26], [27]. Since their first introduction to the photovoltaic community a decade ago, perovskite solar cells have rapidly advanced from 3.9 % to 25.2 % efficiency, according to the efficiency tables published by the National Renewable Energy Laboratory (NREL) [28]. This rapid improvement in the efficiency is partly due to the faster development cycles of perovskite photovoltaics (traditional Si took a few decades to achieve the same performance levels) [29]. Perovskite photovoltaics have the potential to be manufactured at low costs as they do not require high-temperature processing and high-purity materials unlike traditional PV technologies [30]. Additionally, perovskite photovoltaics are solution processable and printable [31], making them more appealing as integrated energy harvesters in printable wireless sensors. Additionally, the color of perovskite PV cells can be customized to align with different brand colors (logos), making them appealing for tagging consumer products. Moreover, the chemical diversity of perovskite PV enables performance optimization for different lighting conditions,

---

[2] https://www.emmicroelectronic.com/sites/default/files/products/datasheets/4325-ds_0.pdf

[3] http://www.farsens.com/wp-content/uploads/2017/12/DS-ROCKY100-V04.pdf

especially artificial lighting (LED and CFL) in indoor environments [32].

### C. Design of a flexible perovskite PV cell

Perovskite photovoltaic cells consist of multiple thin material layers, each having a specific function. The perovskite absorber is the active layer of the photovoltaic cell that harvests incident photons. In addition to the perovskite layer, a perovskite PV cell requires additional layers made of different materials such as Tin (IV) oxide, Titanium (IV) oxide, spiro-OMeTAD, and Gold, which work as charge transport layers (electron transport layer, ETL, and hole transport materials, HTM) and electrodes for efficient flow of current in and out of the device. Comparing to the more established fabrication protocols of perovskite PV on glass, challenges remain to optimize the adhesion between the plastic substrates and the ETL layer to achieve high performance. In this study, we developed a fabrication recipe for the ETL layer consisting of a combination of Tin (IV) Oxide, Titanium (IV) Oxide and Aluminum Chloride Hexahydrate ($AlCl_3.6H_2O$), which allows the improved morphological cover of the ETL layers atop of the plastic substrates (more details of the experimental procedures of cell fabrication is shown in the appendix). To ensure low-temperature manufacturability to reduce costs, all the device fabrication procedures are restricted to under 150 °C. Figure 3 (a) shows the lass substrate, ITO-coated plastic film, perovskite absorber, charge transport layers, and Au top electrode. As in our previous publications, we use $Rb_{0.01}Cs_{0.05}(FA_{0.83}MA_{0.17})_{0.94}Pb(I_{0.83}Br_{0.17})_3$ as the perovskite absorber (bandgap of 1.63 eV) [33]. The proportions of individual chemicals in the perovskite solution are chosen to maximize the material stability, increase efficiency, and increase voltage output. The standard perovskite PV are made on glass and they are not flexible due to the rigidity of the glass. However, perovskite photovoltaics are thin-film cells and they can be made flexible by fabricating on a flexible substrate. Figure 3 (b) shows an image of a fabricated flexible perovskite PV device alongside a standard perovskite PV device on glass. Each fabricated device consists of four individual cells, identifiable by 4 rectangular gold electrodes. The individual cells can be separated from the device to create mini-modules in series or parallel configurations to meet a required current-voltage demand.

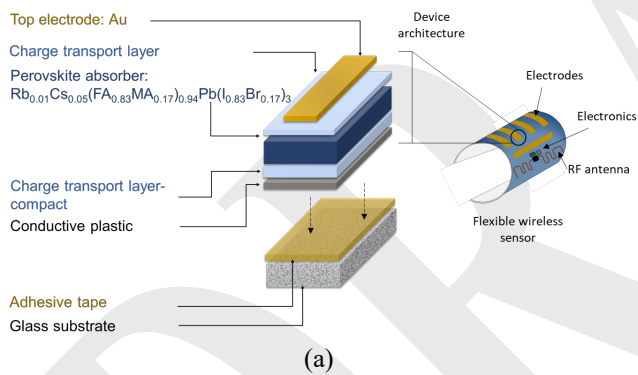

(a)

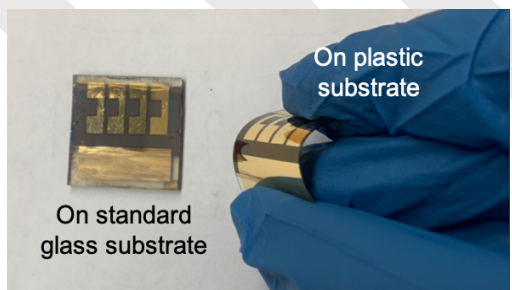

(b)

Figure 3: Perovskite photovoltaic device architecture showing perovskite absorber, charge transport layer, plastic substrate, glass scaffold, and gold electrodes (a); and prototype of a flexible perovskite PV device alongside a standard perovskite PV cell on a glass substrate (b).

### III. TESTING AND EVALUATION OF FLEXIBLE PV

We screened a batch of 30 flexible perovskite PV cells, fabricated in the MIT PV lab, to evaluate their performance: the device's PV conversion efficiency and voltage at the maximum power point. Figure 4 (b) and (c) show the efficiency and voltage measurements in the batch. The best flexible perovskite

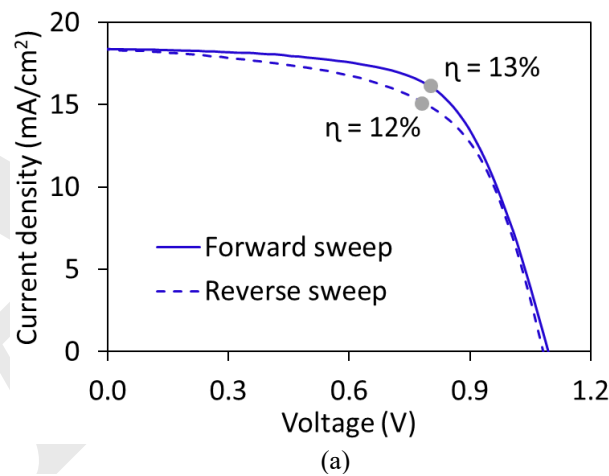

(a)

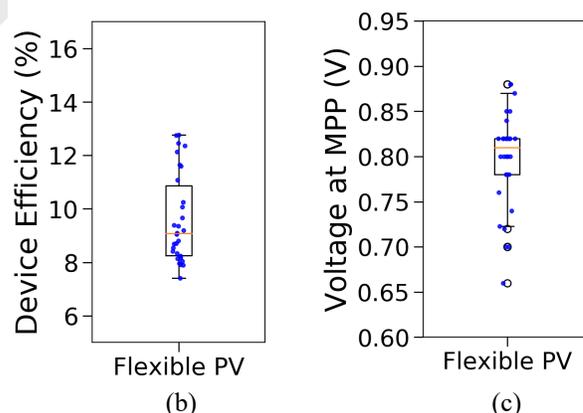

(b)            (c)

Figure 4: Current-voltage relationship in forward and reverse sweep modes and calculated efficiencies (η) at maximum power points (a); measured efficiencies (b) and voltage at maximum power point (c) in a batch of 30 flexible perovskite PV cells [1].

PV cell in the batch has an efficiency of 13%, which is a good efficiency for initial prototypes. Current – voltage relation in the best cell is shown in the Figure 4 (a). The voltage at the maximum power point in the best cell is 0.88V. Fabricating perovskite PV on plastics is a relatively newer process compared to fabricating on glass. This study focuses on demonstrating a proof-of-concept showcase of integrating flexible perovskite PV and RFID devices into a self-powered sensor system, further optimization of the flexible perovskite fabrication process can be obtained through PV device engineering to increase the PV efficiency, similar to developing any other thin-film process. The standard fabrication recipe and process is incrementally improved over a few years to yield such high efficiencies. We are optimistic that the efficiency of flexible perovskite PV will also further increase in the future as the materials and fabrication process are optimized.

The mechanical durability over a range of bending radii is another important metric to evaluate the performance of flexible PV. We measured the cell's efficiency, while bending the cells on a test jig, as a proxy for mechanical durability. The efficiency must be normalized with respect to the efficiency in a flat configuration to estimate the percentage drop due to bending. We used a 3D printed substrate (see Figure 5 a), with curvatures ranging from 200 mm to 5 mm, as a test platform. The results show that the flexible PV cells are durable without any significant efficiency loss due to bending up to 20 mm radius (see Figure 5b). The cell efficiency dropped by 20% due to further bending up to 5 mm. Results are consistent with previous studies [34] showing a similar drop. A 20 % loss is not a significant concern for RFID applications because the loss can be easily compensated by slightly increasing (remember RFID power consumption is in the order of a few µW) the corresponding illumination area of the PV cell. Moreover, 5 mm radius is an aggressive test condition, and the real surfaces of target consumer products, indoor IoT devices, asset trackers, and packages in warehouses are likely to have larger radii of curvatures. Therefore, perovskite PV shows a strong potential to serve as a mechanically flexible energy harvester and conform to a variety of shapes of target objects.[1]

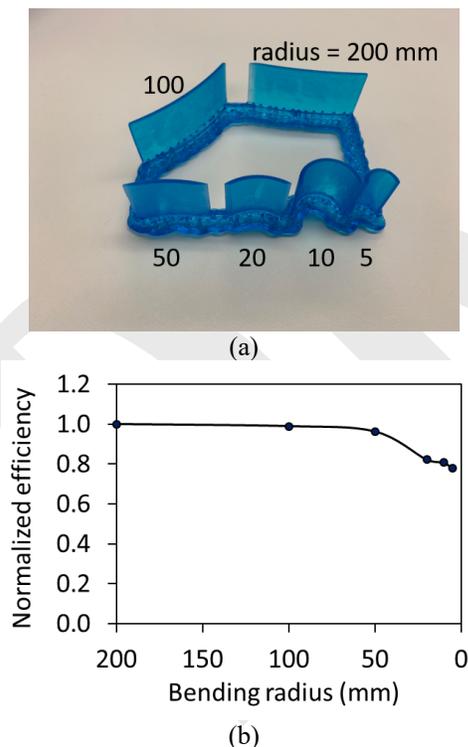

(a)

(b)

Figure 5: 3D printed bending jigs (a); and a plot showing the change in efficiency due to bending (b) [1].

IV. APPLICATION EXPLORATION

Providing extra power by harvesting ambient light using inexpensive flexible PV takes battery-less RFID-based sensors to a higher level in many industry-relevant applications. By increasing the energy available to the RFID tags, we create smart labels that not only provide identification information, but also communicate contextual state information such as temperature in supply chains, product orientation for warehouse automation, and activity recognition in smart spaces. In this section, we present how we built self-powered wireless sensors using flexible perovskite PV and RFID.

*A. Wireless temperature sensor*

Often there is a need to determine storage temperature to maintain product quality of perishable goods in food supply chain. Ensuring the right product storage conditions by temperature monitoring can improve the efficiency of transporting perishable goods. For example, maintaining appropriate temperature for milk jugs, by real-time monitoring using low-cost wireless sensors, can prevent wastage in transit from farm to table. If the storage temperature goes out of range even for a short duration, perishable products like milk are put at risk of spoiling (one in six pints of milk is thrown away each year, according to a study[4]). By developing more advanced sensor labels with time-temperature indicators, consumers can even ensure that the products were never stored in negatively impacting storage conditions [35], [36]. Figure 6 shows a prototype of a wireless temperature sensor using an EM4325 RFID IC powered with flexible perovskite PV. As both the perovskite PV module and the RFID antenna are thin and flexible, the whole sensor can take the curvature of the target object. Moreover, only a small illumination area (in theory, < 1 mm$^2$ at 13% efficiency under 1 Sun illumination) of the perovskite photovoltaics is required to provide the required power (around 15 µW) to the RFID tags with on-board temperature sensors. Electronic circuit diagram to replicate the temperature sensor can be found in our earlier publication on perovskites on glass substrates [8]. If low-power ICs with different embedded sensors (*e.g.* moisture sensor) become

---

[4] https://www.theguardian.com/environment/2018/nov/28/one-in-six-pints-of-milk-thrown-away-each-year-study-shows?CMP=twt_gu&__twitter_impression=true

available, flexible perovskite PV can potentially power those sensors as well.

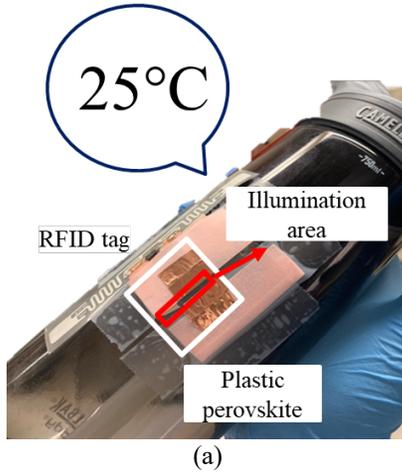

(a)

Figure 6: Prototype of a smart label with flexible perovskite photovoltaic-powered wireless temperature sensor attached to the curved surface of a bottle.

### B. Wireless orientation sensor

Data from inexpensive sensors can also improve robotic automation in distribution warehouses of e-commerce businesses. Robotic automation systems currently rely on RGB and depth data from the cameras to estimate object orientations and 3D poses. These 3D digital poses are then used to estimate pick points for robotic arm to grasp the objects. However, products, especially in distribution warehouses, come in different colors, packing, and textures that makes relying on solely vision-based systems challenging [37]–[41]. Moreover, training 3D pose estimation models for automatic estimation also requires ground truth data from millions (a scale typical in e-commerce distribution) of objects that come in different shapes and sizes. Flexible perovskite-powered RFID-based sensors, with their low-cost, optical tunability (for operation under artificial lighting conditions), and mechanical flexibility, can meet the need for cheap sensor data to augment RGB and depth data from cameras.

Figure 7 (a) illustrates how orientation data acquisition setup can be integrated in a warehouse automation scenario. Products,

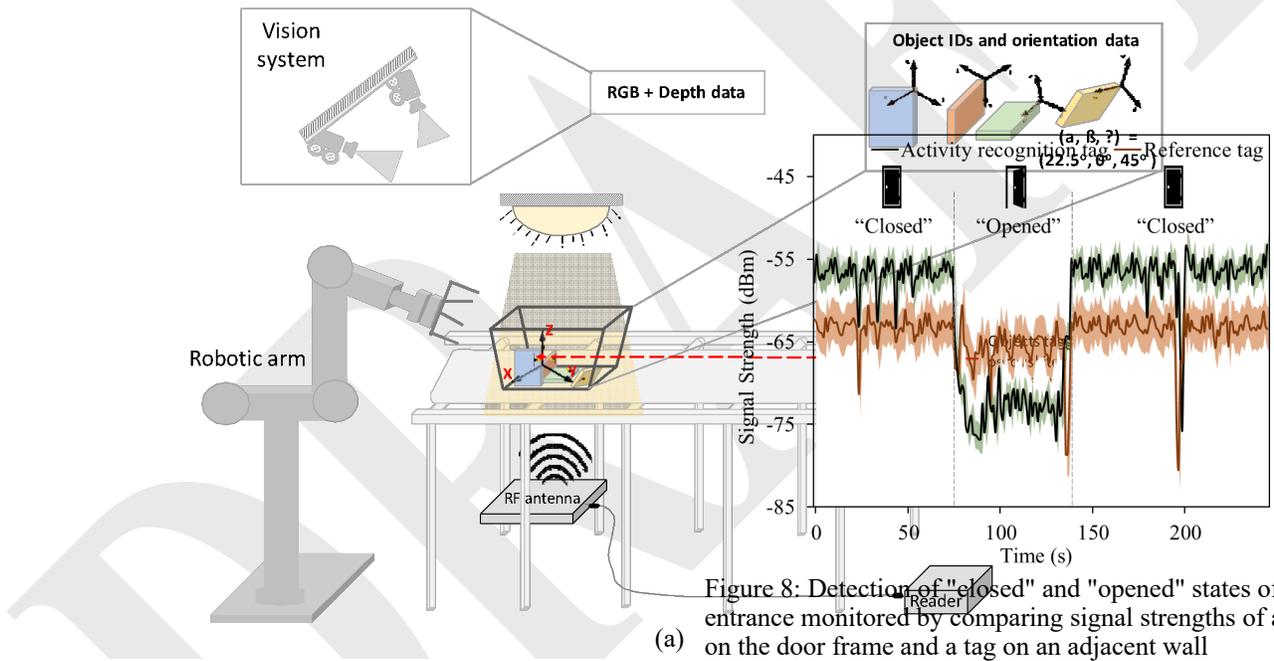

(a)

Figure 8: Detection of "closed" and "opened" states of an entrance monitored by comparing signal strengths of a tag on the door frame and a tag on an adjacent wall

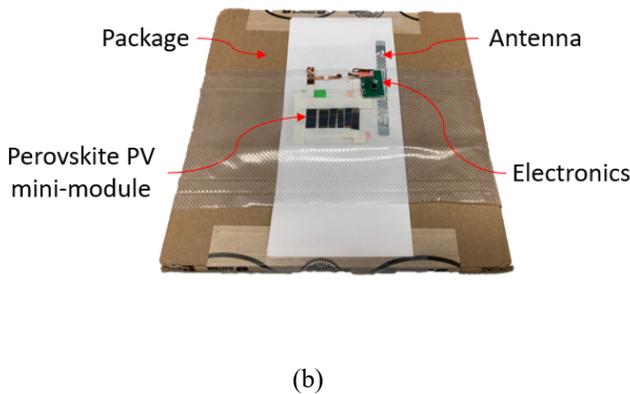

(b)

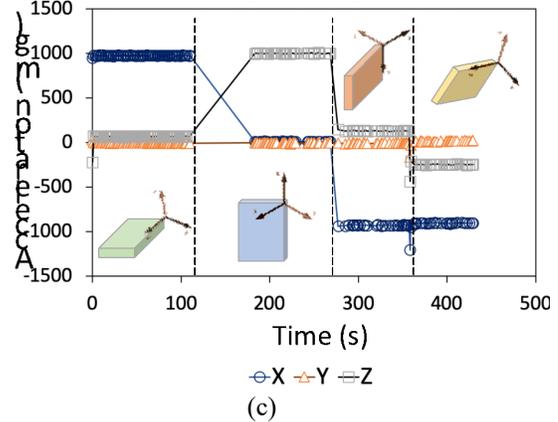

(c)

Figure 7: Illustration of an orientation-acquisition setup using flexible PVID tags on consumer products in a warehouse scenario (a); a prototype of the sensor attached to a product package (b); and a plot showing acceleration data mapped to the orientation of the objects (c)

in different orientations, in bins are carried on conveyors to the robot stations. We can setup an intermediate light curtain to shine light onto the perovskite-powered orientation tags to power them up and acquire orientation data. Figure 7 (b) is a prototype of a perovskite PV-powered RFID orientation sensor attached to an object. The prototype is constructed by modifying a Farsen's sensor [42], and it consists of an orientation sensor [43] and a microcontroller [44]. Estimated peak power consumption for the whole device is around 350 µW. Also, the required voltage output is greater than 3 V from the perovskite PV modules. We use a mini-module with 6 cells in series to meet the required power demand. Figure 7 (c) shows the acceleration data acquired from the orientation sensor that can be mapped to the orientation of the product (assuming the measurements are taken at stationary positions). This orientation data obtained using low-cost wireless sensors can help in two folds: 1) data fusion with the data from cameras to improve the speed and accuracy of 3D pose estimation; and 2) as ground truth data to train deep learning modules in computer vision to build and implement object segmentation and pose estimation models.

### C. Activity recognition sensors

Perovskite-powered RFID tags also have potential as low-cost, long-range wireless activity recognition sensors (see Figure 8). These low-cost, easy to deploy, and battery-less tags enable us to easily glean data from everyday objects around us. This data can be used to for activity recognition or monitoring in variety of applications. In this study, we demonstrate recognizing opened/closed state of an entrance in our academic building by tagging the door frame with perovskite-powered RFID tags. The technique we propose here relies on detecting these events by comparing the signal strength of two tags: an activity recognition tag and a reference tag. For example, as illustrated in Figure 8, activity recognition tag is directly attached to the door frame so that it is more susceptible to when the door is opened or closed, and a reference tag is attached to an adjacent wall for reference signals. When the door is opened/closed the signal from the activity recognition tag significantly deviates from the signal from reference tag indicating that a door opened/closed event has occurred. Figure 9 shows an average signal strength along with their variabilities for both activity recognition and reference tags. By comparing the signals over time, different windows of activities can be detected. This concept can be further extended to more sophisticated events such as monitoring vibrations, motion, etc. by building appropriate data processing models.

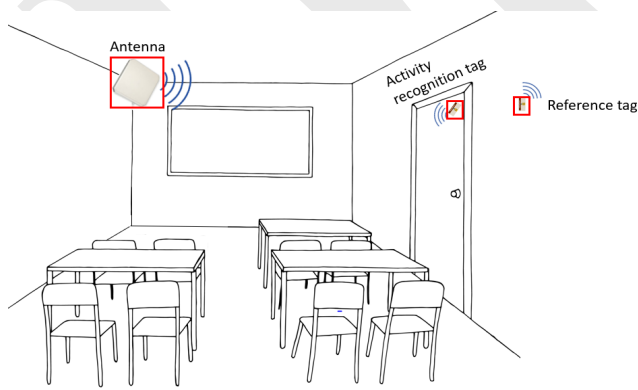

Figure 9: Illustration of activity recognition system setup in a classroom

## V. FUTURE WORK

Although perovskites already show a great promise as energy harvesters for the IoT, there is a significant scope for future research to further improve perovskite PV. This section discusses a few avenues for future work in the context of powering IoT devices.

### A. Stability

For IoT applications, we require devices with a few days to years of life. Faster degradation rates of perovskite PV due to exposure to moisture, heat, air, and light is currently a barrier to deploy these cells in real world, for long durations. This emphasizes a need to improve the environmental stability of perovskite PV cells. Development directions include developing better packaging and degradation-resistant materials. There is a need to explore new structural and encapsulation strategies to improve the stability, for examples, new approaches (e.g., encapsulation techniques) are being developed as described in [45], [46].

### B. Indoor toxicity of perovskite materials containing lead

The presence of lead in the current high-efficiency perovskite PV materials is a concern for deployment in indoor applications. Currently, lead-free perovskite materials have lower efficiencies, and the search for high-efficient lead-free perovskite PV materials is still ongoing [47], [48]. There is a need to explore new doping materials and strategies to improve the efficiency of lead-free perovskite photovoltaics.

### C. Powering other RF backscatter technologies

This study presents flexible perovskite PV in combination with RFID tags. However, perovskite PV can also meet the power demand of other emerging low-power RF backscatter technologies such as LoRa backscatter [49]. and WiFi backscatter [50]. We need to investigate the sizing of PV cells and the required current draw to power such newer technologies.

## VI. CONCLUSION

We presented an emerging perovskite-based thin-film photovoltaic technology that is able to create mechanically flexible, optically tunable, high-performance, and low-cost energy harvesters for the IoT. Combining with the low-cost, low-power RFID technology, we can create wireless sensors to augment billions of consumer products to create new experiences, improve supply chain visibility, improve warehouse automation, and improve product quality.

Using prototypes of wireless temperature, orientation and activity recognition sensors powered with flexible perovskite PV, we demonstrate how we can create conformal self-powered wireless sensors for tagging consumer products. Data acquired from these inexpensive sensor labels can improve processes in supply chains, warehouse automation, and smart spaces applications. Our evaluation of the prototypes shows that the best cells are 13% efficient at 0.88 V output at maximum power point. The results from the flexibility tests show that cells are durable without any significant efficiency loss due to bending up to 20 mm radius. A mini-module of cells is sufficient to successfully power the RFID tags and auxiliary electronics needing 10-350 µW of power. Primary advantages of using this emerging photovoltaic technology — over traditional PV that the IoT world is more familiar with — are the manufacturing synergies with RFID tags and the ability to tune perovskite PV for various operating conditions levering the enormous chemical diversity of these materials.

APPENDIX

*A. Fabrication process*

Flexible PET (from Sigma Aldrich) coated with Indium Tin Oxide (ITO) was mounted on the glass scaffold with a double sided Kapton under the edges. ITO coated plastic (10 Ω/sq) was gently scrubbed with Kleenex soaked in ethanol and Isopropylalcohol (Macron Chemicals). UV ozone treatment for 15 min was implemented to further clean the surface. For the electro-transport layer (ETL), a combination of Tin (IV) Oxide, Titanium (IV) Oxide and Aluminum Chloride Hexahydrate ($AlCl_3.6H_2O$) layers was deposited. An aqueous solution of 15% $SnO_2$ (Alfa Aesar) was diluted to 3% using DI water for $SnO_2$ layer. For the $TiO_2$ layer, 19.0 wt. % anatase Titania paste (Greatcell Solar) was diluted in ethanol in a 1:7 weight ratio. $AlCl_3.6H_2O$ (Sigma-Aldrich) was dissolved in ethanol in 0.5% weight ratio as solution for $Al2O3$ layer. The same solution was as a dopant in cases where Al-doped layers were used. All layers were separately spin-coated at 4500 rpm for 30 sec. followed by annealing at 150°C for 1 hour before depositing additional layers atop. The perovskite precursor solution was prepared, as in our previous publications, by mixing FAI (1 M, Dyenamo), $PbI_2$ (1.1 M, TCI), MABr (0.2 M, Dyenamo) and $PbBr_2$ (0.22 M, TCI) in a 9:1 (v:v) mixture of anhydrous DMF:DMSO (Sigma Aldrich). CsI (Sigma Aldrich) in DMSO solvent and RbI (Sigma Aldrich) in 5:1 (v:v) DMF:DMSO solvent were respectively added in a 5:1:95 volume ratio to form CsI:RbI:perovskite solution. Using a two-step spin coating program (10 s at 1000 rpm, 20 s at 6000 rpm), the solution was spin-coated on the UV-cleaned substrate coated with ETL layers. An 150 µL antisolvent of chlorobenzene was added during the second step of spincoating. The films were then annealed at 130 °C for 20 min. We used Spiro-OMeTAD (2,2′7,7′-tetrakis-(N,N-di-p-methoxyphenyl amine)-9,9′-spirobifluorene, LumTec LT-S922) as the hole-transport material. Every gram of spiro-OMeTAD was mixed with 227 µL of Li-TFSI (Sigma-Aldrich, 1.8 M in acetonitrile) solution, 394 µL of 4-tert-butylpyridine (Sigma-Aldrich) solution, 98 µL cobalt complex (FK209, Lumtec, 0.25 M tris(2-(1H-pyrazol-1-yl)-4-tertbutylpyridine) cobalt(III) tris(bis(trifluoromethylsulfonyl)imide) in acetonitrile) solution, and 10,938 µL of chlorobenzene. 60µL of the mixed spiro solution was spincoated onto the perovskite films at 3000 rpm for 30 s. One side of HTM coated sample was wiped near the edge with 2-Methoxyethanol to open contact to the ETL layer. Finally, metallization was performed with via thermal deposition to deposit 100 nm gold electrode.